\documentclass[preprint,12pt]{elsarticle}

\usepackage{amssymb}
\usepackage{amsmath}
\usepackage{caption}
\usepackage{subcaption}
\usepackage[colorlinks=true,
            linkcolor=blue,
            citecolor=blue,
            urlcolor=blue]{hyperref}

\journal{Journal of Subatomic Particles and Cosmology}

\begin{document}

\begin{frontmatter}



\title{Probes of lepton flavor symmetry and violation with top quarks in ATLAS and CMS\tnoteref{copyright}} 
\tnotetext[copyright]{Copyright 2026 CERN for the benefit of the ATLAS and CMS Collaborations. CC-BY-4.0 license.}

\author{Normunds Ralfs Strautnieks, for the ATLAS and CMS collaborations} 
\ead{normunds.ralfs.strautnieks@cern.ch}
\affiliation{organization={University of Latvia},
            city={Riga},
            country={Latvia}}

\begin{abstract}
Measurements of lepton flavor universality (LFU) and searches for charged lepton flavor violation (cLFV) are one of the most straightforward ways of searching for Beyond Standard Model (BSM) physics. The numerous production of top quark pairs at the LHC allows for high precision measurements of LFU ratios and searches for cLFV signatures. Results of the recent measurements and searches from the ATLAS \cite{ATLAS_Collab} and CMS \cite{CMS_Collab} collaborations using the Run2 dataset with an integrated luminosity of up to 139 $\mathrm{fb}^{-1}$ at center of mass energy of 13 TeV are presented.
\end{abstract}


\begin{keyword}
LFU \sep cLFV \sep LHC \sep ATLAS \sep CMS

\end{keyword}
\end{frontmatter}

\section{Introduction}
\label{sec1}
The Standard Model (SM) of elementary particles is one of the most precise and predictive physics theories of modern science, from which the predictions are still tested and verified today. Despite it being a successful theoretical framework, it is an incomplete theory. Because of this fact a search for BSM physics is an increasingly active area of scientific research. There are many avenues in which BSM effects could potentially be observed and one group of them are LFU measurements and searches for cLFV decays. 
For the LFU measurements, the focus is mainly on B meson and leptonic decays of the W and Z bosons. For measurements involving W boson, the top quark is particularly important as more than 99\% of top quarks decay into a W boson. Meanwhile, the cLFV decays are searched for in Z boson, Higgs and top quark decays into multiple leptons of different flavors. If an observation is made that LFU ratios deviate from unity or any cLFV decays are observed, that would be a clear indication of BSM physics and could imply the existence of new elementary particles like leptoquarks, Higgs doublets and additional Z' and W' bosons among many others.

\section{Lepton Flavor Universality}
\label{sec2}
\begin{figure}[htb]
    \centering
    \begin{subfigure}[b]{0.40\textwidth}
        \centering
        \includegraphics[width=\textwidth]{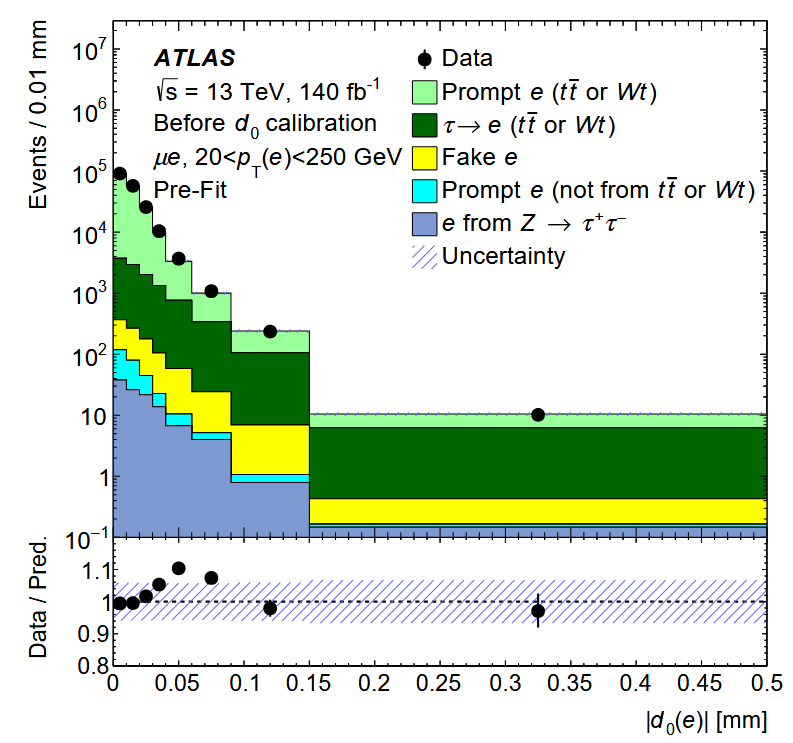}
        \caption{}
        \label{fig:Ele_dxy_before}
    \end{subfigure}
    \hfill
    \begin{subfigure}[b]{0.40\textwidth}
        \centering
        \includegraphics[width=\textwidth]{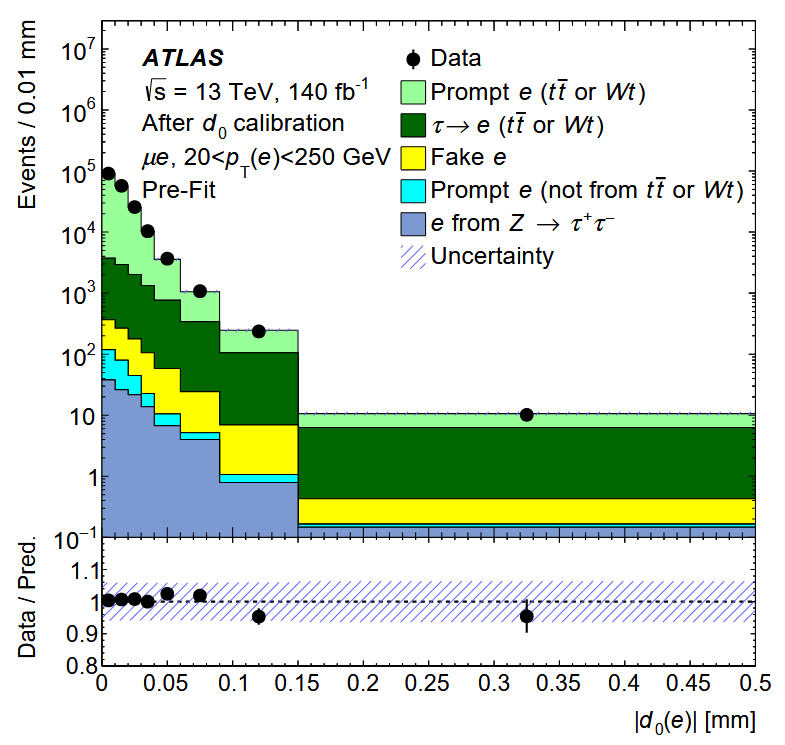}
        \caption{}
        \label{fig:Ele_dxy_after}
    \end{subfigure}
    \caption{The electron transverse displacement distributions before (a) and after (b) the transverse displacement's calibration procedure is performed. Figure taken from \cite{ATLAS_TauE}.}
    \label{fig:Lepton_nominal_dxy}
\end{figure}

\begin{figure}[htb]
    \centering
    \begin{subfigure}[b]{0.42\textwidth}
        \centering
        \includegraphics[width=\textwidth]{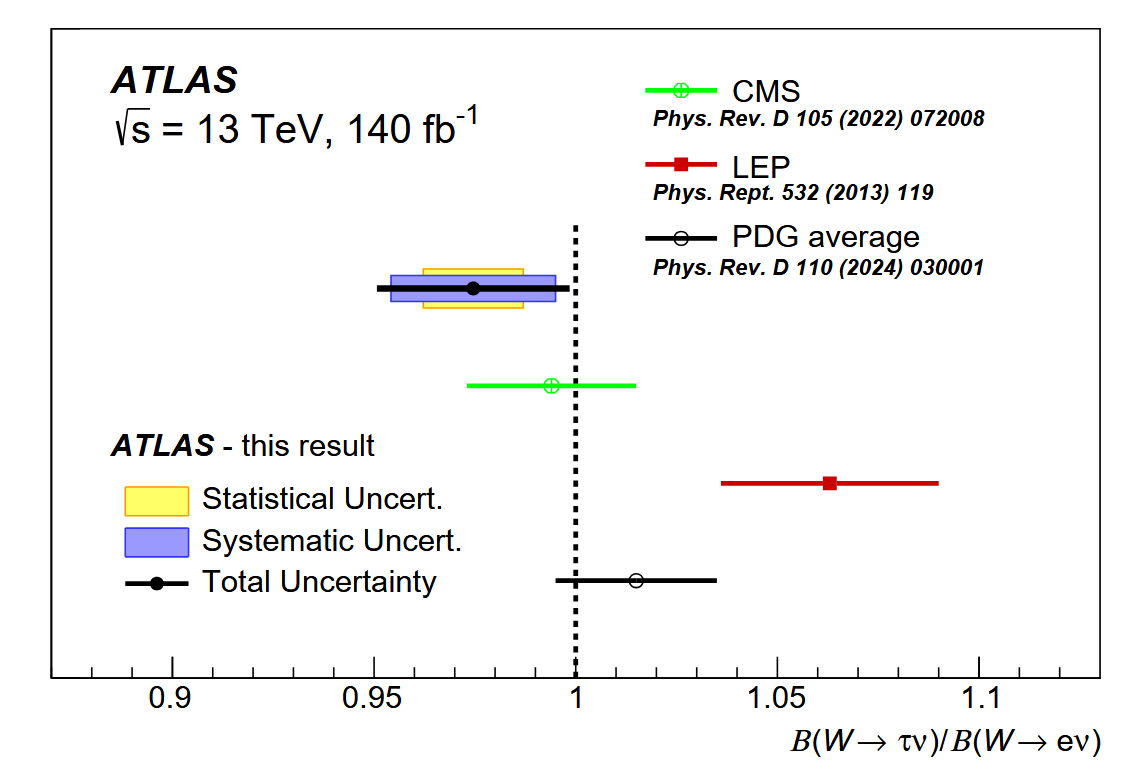}
        \caption{}
        \label{fig:R_tau_e_ATLAS}
    \end{subfigure}
    \hfill
    \begin{subfigure}[b]{0.42\textwidth}
        \centering
        \includegraphics[width=\textwidth]{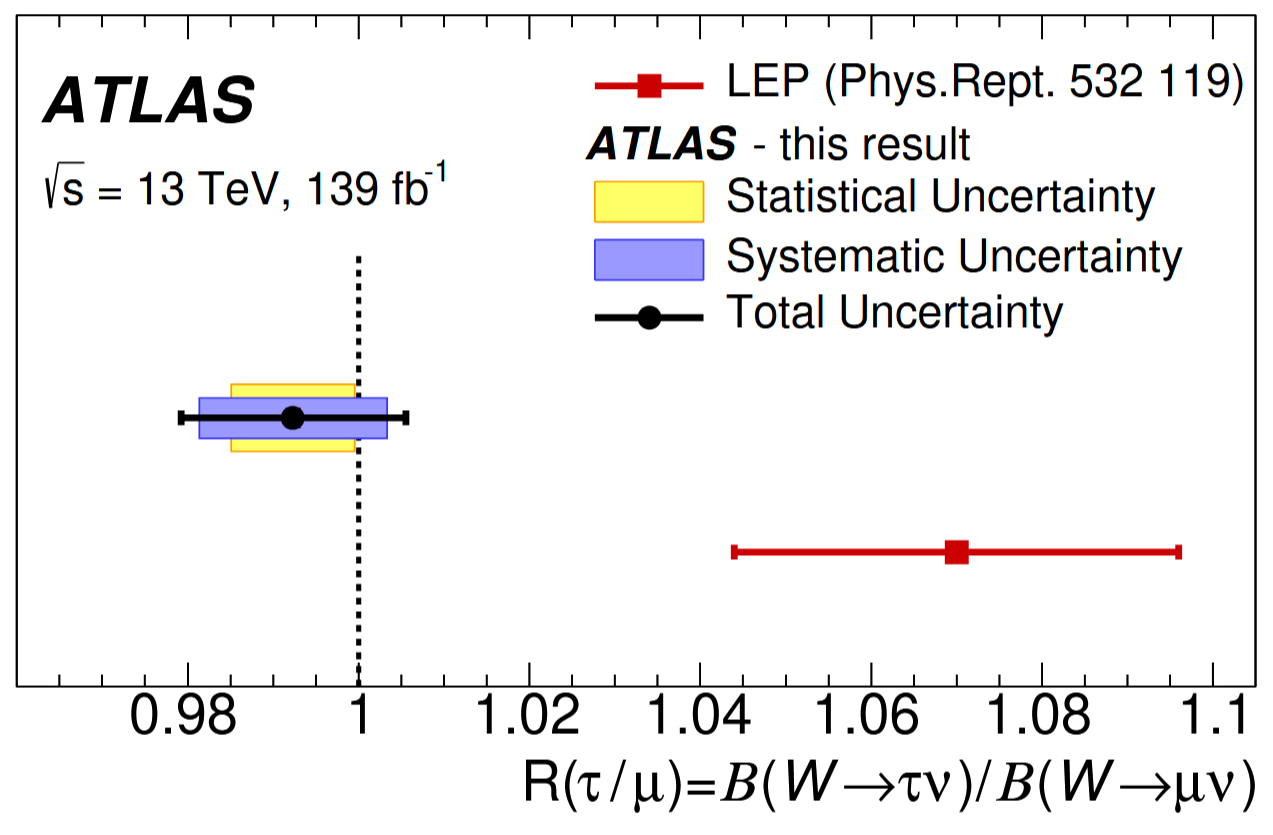}
        \caption{}
        \label{fig:R_tau_mu_ATLAS}
    \end{subfigure}
    \caption{Final results for the LFU ratio R($\tau/e$) \cite{ATLAS_TauE} (a) and R($\tau/\mu$) \cite{ATLAS_TauMu} (b) measurements by the ATLAS collaboration using the LHC Run2 dataset.}
    \label{fig:LFU_ATLAS}
\end{figure}

The condition of LFU is an axiom of SM that postulates that the gauge boson couplings are independent of the lepton flavor and it can be directly tested in Z and W boson decays. A recent measurement from the ATLAS collaboration has measured the LFU ratio R($\tau/e$) using the full LHC Run2 dataset of 139 $\mathrm{fb}^{-1}$ at 13 TeV center of mass energy. In this analysis the R($\tau/e$) is measured as branching ratio (BR) shown in equation~\ref{eq:LFU_ratio_eq}, where the W bosons needed for the measurement are acquired from the top quark pair decay process.
\begin{equation}
\label{eq:LFU_ratio_eq}
R_{\tau/e} = \frac{B(W\rightarrow \tau\nu)}{B(W\rightarrow e\nu)}
\end{equation}
For this analysis the $\tau$ leptons are selected in their leptonic decay channel where they decay to electrons. Because of this approach, there needs to be a way to identify electrons stemming directly from a W decay (prompt) vs the ones from a $\tau$ decay (non-prompt). This is achieved by exploiting the differences in the electron $p_T$ and transverse displacement ($d_0$) distributions between prompt and non-prompt electrons, so that $R_{\tau/e}$ can be extracted via a 2-D maximum likelihood fit. 
There is an intrinsic difference between the MC and data resolutions of electron $d_0$, which is calibrated using data-driven transverse displacement templates in 39 $p_T/\eta$ bins to describe the distributions with as high accuracy as possible.
A comparison of electron transverse displacement before and after calibration is shown in Fig.~\ref{fig:Lepton_nominal_dxy}. After the calibrations are performed, the final LFU ratio is extracted and the results can be seen in Fig.~\ref{fig:R_tau_e_ATLAS} together with measurements from LEP, CMS, and the PDG average. Previously, a measurement of R($\tau/\mu$) was carried out by ATLAS collaboration using a similar technique, for which the result is shown in Fig.~\ref{fig:R_tau_mu_ATLAS}. The results of both measurements are consistent with the SM prediction. More details can be found in \cite{ATLAS_TauE} and \cite{ATLAS_TauMu}.

\begin{figure}[htb]
    \centering
    \begin{subfigure}[b]{0.40\textwidth}
        \centering
        \includegraphics[width=\textwidth]{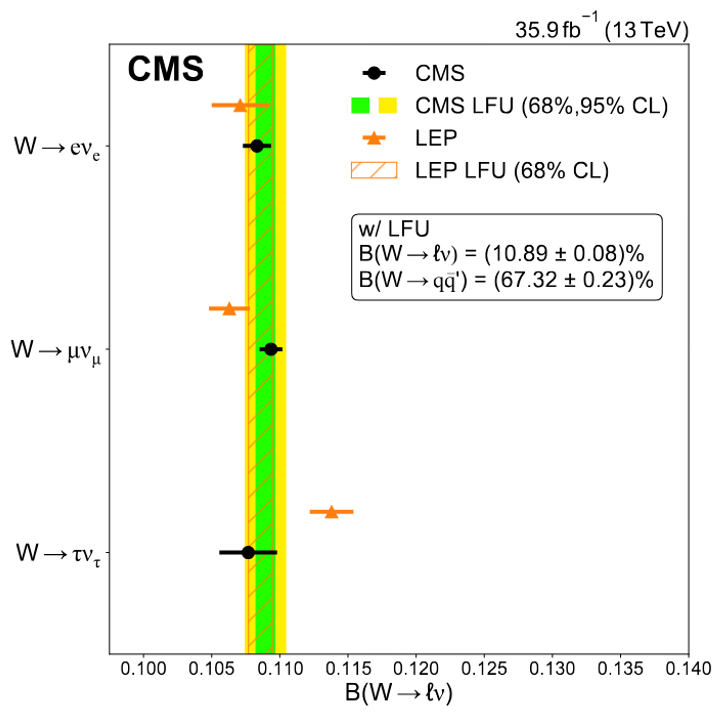}
        \caption{}
        \label{fig:CMS_W_BR}
    \end{subfigure}
    \hfill
    \begin{subfigure}[b]{0.40\textwidth}
        \centering
        \includegraphics[width=\textwidth]{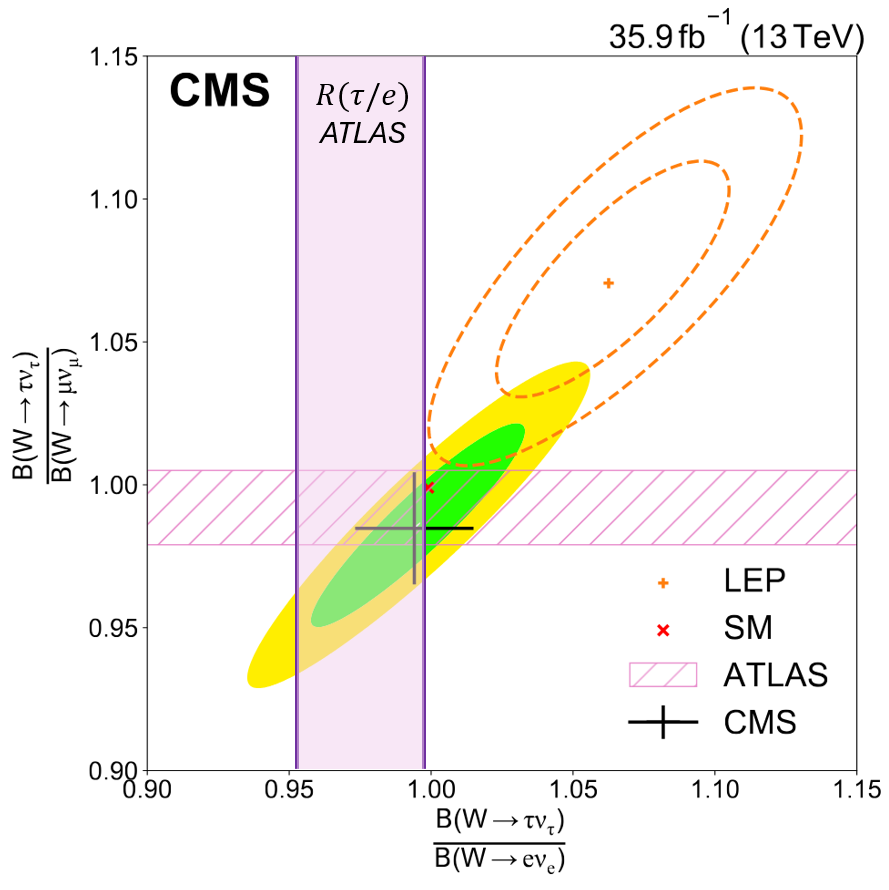}
        \caption{}
        \label{fig:CMS_W_LFU}
    \end{subfigure}
    \caption{The measured W boson leptonic BR results (a) and the correspondent correlated LFU ratio results with their comparison to the values measured by the ATLAS and LEP collaborations (b). Figure taken from \cite{CMS_LFU}.}
    \label{fig:LFU_CMS}
\end{figure}

An alternative approach for the W boson branching fraction and the corresponding LFU ratio measurement is carried out by the CMS collaboration. Here, a binned likelihood fit is performed in $p_T$ bins and 30 event categories defined by lepton multiplicity and their flavor, the number of total jets, and the number of b-tagged jets. This categorization split helps to better describe the signal and background contributions, while using all reconstructible W boson and $\tau$ lepton decay modes. For this measurement most of the signal events arise from a top pair decay process with a small contribution of WW boson production. The analysis is performed using the 2016 dataset of 35.9~$\mathrm{fb}^{-1}$ at center of mass energy of 13 TeV. The final results are summarized in Fig.~\ref{fig:LFU_CMS}, where the leptonic W BRs are shown to be in agreement with the SM result with decreased uncertainties compared to the previous LEP measurements. Additionally, the 2D contours of the LFU ratios R($\tau/\mu$) and R($\tau/e$) are shown, where they are compared to the results reported by the ATLAS collaboration and the previous LEP measurement. More details on the measurement can be found in \cite{CMS_LFU}.

\section{Charged Lepton Flavor Violation}
\label{sec3}
\begin{figure}[htb]
    \centering
    \begin{subfigure}[b]{0.42\textwidth}
        \centering
        \includegraphics[width=\textwidth]{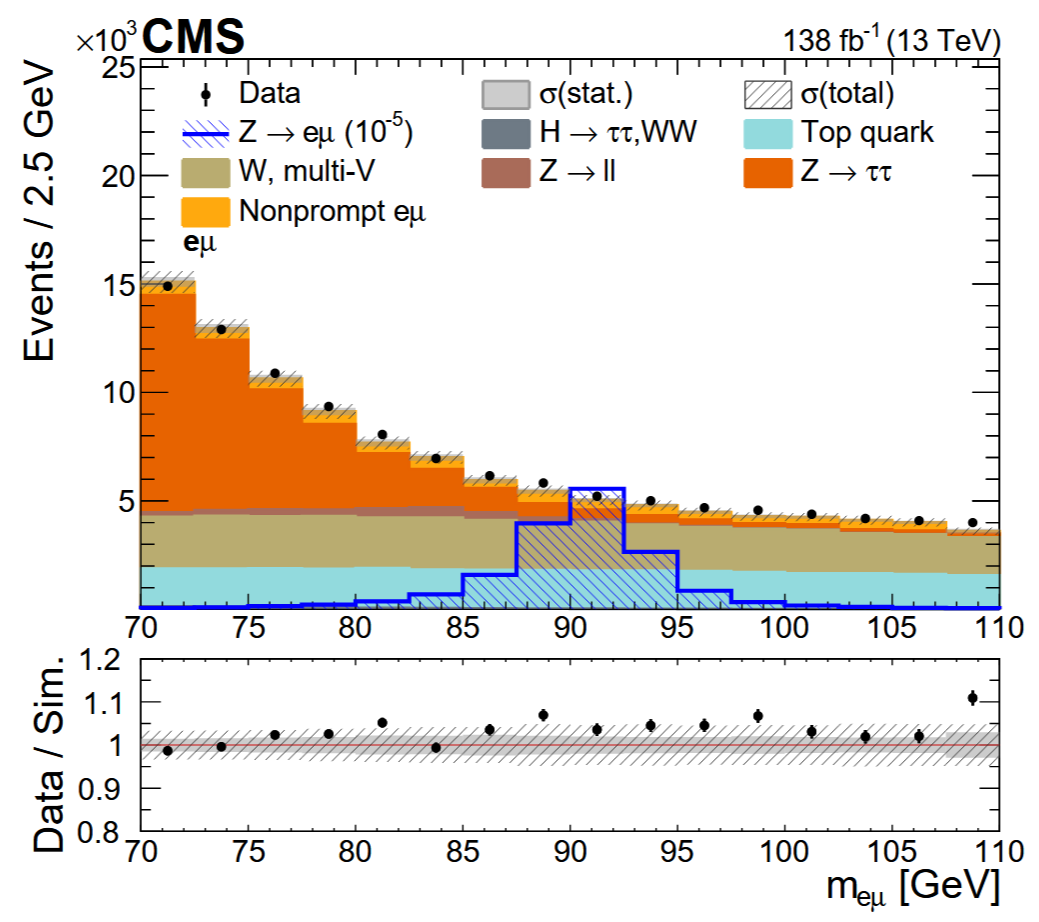}
        \caption{}
        \label{fig:emu_InvM}
    \end{subfigure}
    \hfill
    \begin{subfigure}[b]{0.45\textwidth}
        \centering
        \includegraphics[width=\textwidth]{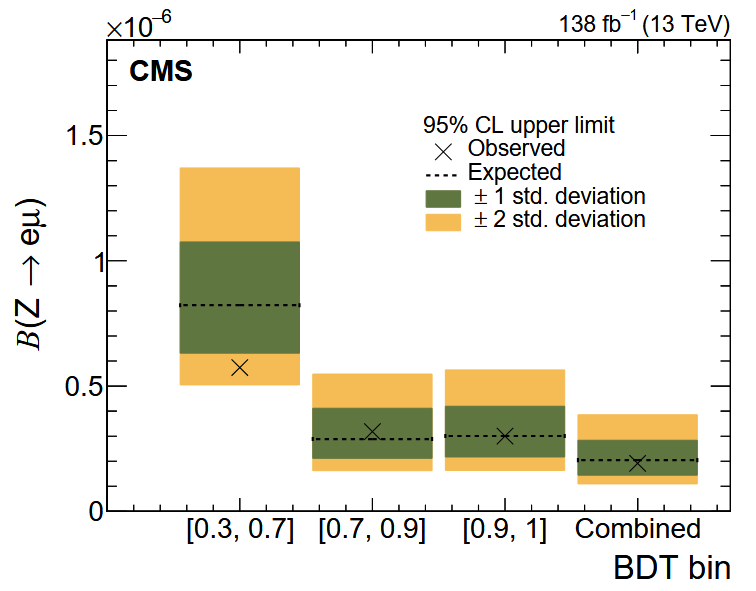}
        \caption{}
        \label{fig:cLFV_emu}
    \end{subfigure}
    \caption{The invariant mass distribution with an overlaid $Z\rightarrow e\mu$ signal if the BR for the the cLFV decay was $10^{-5}$ (a) and the corresponding cLFV decay branching ratio results for BR($Z\rightarrow e\mu$) (b). Figure taken from \cite{CMS_Z_cLFV}.}
    \label{fig:cLFV_CMS_BR_lightLep}
\end{figure}

The cLFV decays are forbidden in the SM at the tree diagram level and are only allowed through higher order neutrino mixing loop diagrams. This puts the expected SM BR to a level of around $10^{-60}$, which means that they are impossible to observe at any of the LHC experiments. Thus, any cLFV decay observed at the LHC would be a clear signal of BSM physics. A search for these cLFV decays is carried out by the CMS collaboration, where a measurement of $Z\rightarrow e\mu$, $Z\rightarrow e\tau$, $Z\rightarrow \mu\tau$ decays is performed. To search for the cLFV signal, a boosted decision tree (BDT) is trained on signal MC events with background MC contributions coming from top quark pair and WW production. Afterwards, three signal regions, based on the BDT output score, are used in a combined fit for signal and background contributions. The signal is described with a double sided crystal ball function and the background is modeled by the best-fitting function from Chebyshev, exponential, or power law functions. In Fig.~\ref{fig:emu_InvM} the $e\mu$ invariant mass distribution is shown with all the backgrounds and a potential signal signature (scaled to a higher BR for it to be visible), while in Fig.~\ref{fig:cLFV_emu} the extracted BR limit for the $Z\rightarrow e\mu$ decays is presented at $1.9\times10^{-7}$, which is the most stringent limit from the LHC experiments up to date, with the previous measurement coming from the ATLAS collaboration at $2.6\times10^{-7}$. More details on these measurements can be found in \cite{CMS_Z_cLFV} and \cite{ATLAS_Z_cLFV1}.

\begin{figure}[htb]
    \centering
    \begin{subfigure}[b]{0.42\textwidth}
        \centering
        \includegraphics[width=\textwidth]{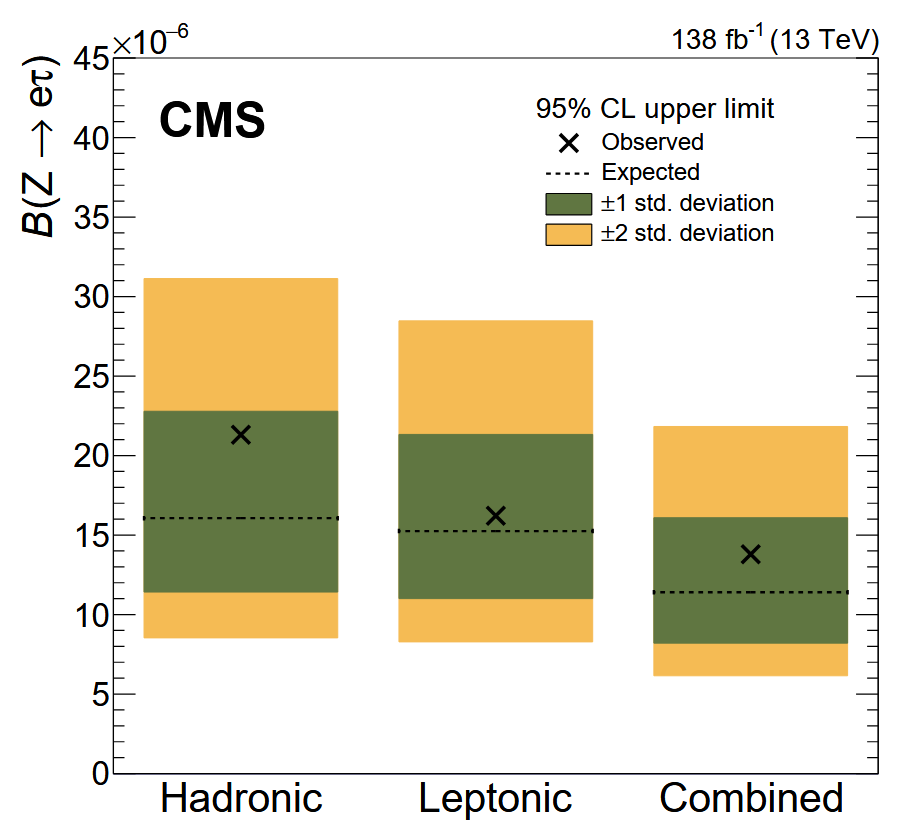}
        \caption{}
        \label{fig:cLFV_BR_tauE}
    \end{subfigure}
    \hfill
    \begin{subfigure}[b]{0.42\textwidth}
        \centering
        \includegraphics[width=\textwidth]{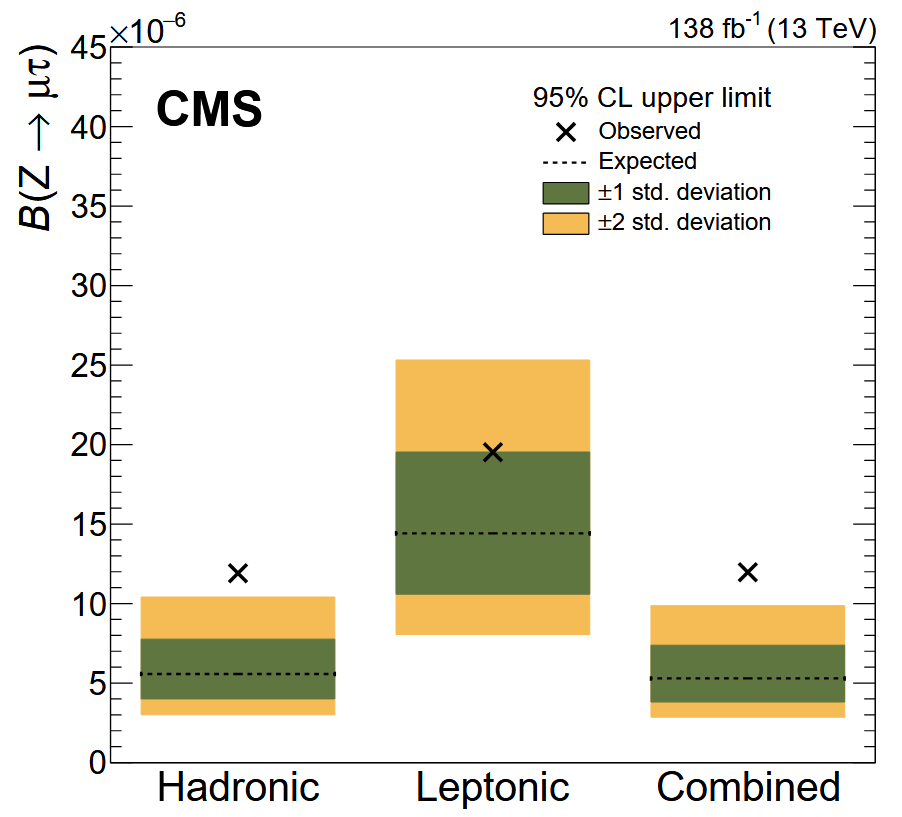}
        \caption{}
        \label{fig:cLFV_BR_tauMu}
    \end{subfigure}
    \caption{The summary results of limits for the cLFV decay branching ratios BR($Z\rightarrow e\tau$) (a) and BR($Z\rightarrow \mu\tau$) (b). Figure taken from \cite{CMS_Z_cLFV}.}
    \label{fig:cLFV_CMS_BR_tau}
\end{figure}

Additional to the $Z\rightarrow e\mu$ limit measurement, the limits on $Z\rightarrow e\tau$ and $Z\rightarrow \mu\tau$ BRs are derived using a slightly modified BDT approach, where instead of fitting the invariant mass distribution, the BDT output distribution itself is fitted. This is done due to the fact that the $e\tau$ and $\mu\tau$ invariant mass distributions are much broader than the $e\mu$ one and provides less sensitivity to the limit extraction. The BR limits for both hadronic and leptonic $\tau$ decay modes are summarized in Fig.~\ref{fig:cLFV_CMS_BR_tau}. The limits for $Z\rightarrow e\tau$ and $Z\rightarrow \mu\tau$ are $13.8\times10^{-6}$ and $12.0\times10^{-6}$, respectively, which are slightly lower than the limits obtained from the same measurement performed by ATLAS collaboration, which set corresponding limits of $7.0\times10^{-6}$ and $7.2\times10^{-6}$. More details on the measurements can be found in \cite{CMS_Z_cLFV} and \cite{ATLAS_Z_cLFV2}.

\section{Conclusions}
\label{sec4}
Both the ATLAS and CMS collaborations have a detailed analysis program for LFU measurements and searches for cLFV decays. As discussed in these proceedings, the most stringent limits are set for finding cLFV decays from Z boson and the highest precision results are acquired for the LFU ratios. It is important to note that the discussed LFU measurements have focused on obtaining the R($\tau/l$) ratios instead of focusing only on R($\mu/e$), which has been the main focus of the B meson physics LFU ratio program. This clearly illustrates that it is possible to do precision physics measurements using $\tau$ leptons. With the presented analysis methods and the Run3 data-taking coming to an end, the prospects for future analysis results from both the ATLAS and CMS collaborations using the combined Run2 and Run3 datasets are exciting. 




\begin{thebibliography}{00}


\bibitem[ATLAS(2008)]{ATLAS_Collab}
  ATLAS Collaboration,
  \textit{The ATLAS Experiment at the CERN Large Hadron Collider},
  2008 JINST 3 S08003,
  (2008),
  doi:10.1088/1748-0221/3/08/S08003.

\bibitem[CMS(2008)]{CMS_Collab}
  CMS Collaboration,
  \textit{The CMS experiment at the CERN LHC},
  2008 JINST 3 S08004,
  (2008),
  doi:10.1088/1748-0221/3/08/S08004.

\bibitem[ATLAS(2025)]{ATLAS_TauE}
  ATLAS Collaboration,
  \textit{Test of lepton flavor universality in W boson decays into electrons and $\tau$ leptons using $pp$ collisions $\sqrt{s}=$ 13 TeV with the ATLAS detector},
  JHEP 05 (2025) 038,
  (2025),
  doi:10.1007/JHEP05(2025)038.

\bibitem[ATLAS(2021)]{ATLAS_TauMu}
  ATLAS Collaboration,
  \textit{Test of the universality of $\tau$ and $\mu$ lepton couplings in W boson decays with the ATLAS detector},
  Nat. Phys. 17, 813–818,
  (2021),
  doi:10.1038/s41567-021-01236-w.

\bibitem[CMS(2022)]{CMS_LFU}
  CMS Collaboration,
  \textit{Precision measurement of the W boson decay branching fractions in proton-proton collisions at $\sqrt{s}=$ 13 TeV},
  Phys. Rev. D 105 (2022) 072008,
  (2022),
  doi:10.1103/PhysRevD.105.072008.

\bibitem[CMS(2025)]{CMS_Z_cLFV}
  CMS Collaboration,
  \textit{Search for charged lepton flavor violating Z and Z' boson decays in proton-proton collisions at $\sqrt{s}=$ 13 TeV},
  CERN-EP-2025-130,
  (2025),
  doi:10.48550/arXiv.2508.07512.

\bibitem[ATLAS(2023)]{ATLAS_Z_cLFV1}
  ATLAS Collaboration,
  \textit{Search for the charged-lepton-flavor-violating decay $Z\rightarrow e\mu$ in $pp$ collisions at $\sqrt{s}=$ 13 TeV with the ATLAS detector},
  Phys. Rev. D 108, 032015,
  (2023),
  doi:10.1103/PhysRevD.108.032015.

\bibitem[ATLAS(2021)2]{ATLAS_Z_cLFV2}
  ATLAS Collaboration,
  \textit{Search for Lepton-Flavor Violation in Z-Boson Decays with $\tau$ Leptons with the ATLAS Detector},
  Phys. Rev. Lett. 127, 271801,
  (2021),
  doi:10.1103/PhysRevLett.127.271801.

\end{thebibliography}
\end{document}